\input harvmac
%
%%%%%%%%%%%%%%%%%%%%%%%%%%%%%%%%%%%%%%%%%%%%%%%%%%%
\rightline{RUNHETC-2001-8}
\Title{
\rightline{hep-th/0103089}}
{\vbox{\centerline{On The BV Formulation}
\vskip 10pt
\centerline{Of Boundary Superstring Field Theory}}}
\bigskip
\centerline{Marcos Mari\~no}
\bigskip
\centerline{\it New High Energy Theory Center}
\centerline{\it Rutgers University}
\centerline{\it Piscataway, NJ 08855-0849, USA}
\centerline{marcosm@physics.rutgers.edu}

\bigskip
\noindent
We propose a Batalin-Vilkovisky (BV) formulation of 
boundary superstring field theory. The superstring field 
action is defined in terms of a closed one-form in the 
space of couplings, and we compute it explicitly for 
exactly solvable tachyon perturbations. We also argue that 
the superstring field action defined in this way is the 
partition function on the disc, in accord with a previous 
proposal.  
\vfill

\Date{March 12, 2001} 

\lref\akbm{I.~Y.~Aref'eva, A.~S.~Koshelev, D.~M.~Belov and P.~B.~Medvedev, 
``Tachyon condensation in cubic superstring field theory,''
hep-th/0011117.}
\lref\bsft{E. Witten, ``On background
independent open string field theory," hep-th/9208027, 
Phys. Rev. {\bf D 46} (1992) 5467.}
\lref\hkm{J. Harvey, D. Kutasov and E. Martinec, ``On the relevance 
of tachyons,'' hep-th/0003101.}
\lref\kmms{D. Kutasov, M. Mari\~no and G. Moore,
``Remarks on tachyon condensation in superstring
field theory," hep-th/0010108.}
\lref\witten{E. Witten, ``Interacting field theory of
open superstrings," Nucl. Phys. {\bf B 276} (1986) 291.}
\lref\fms{D. Friedan, E. Martinec, and S. Shenker,
``Conformal invariance, supersymmetry, and string theory,"
Nucl. Phys. {\bf B 271} (1986) 93.}
\lref\wittwo{E. Witten, ``Some computations in background-independent
off-shell string theory," hep-th/9210065, Phys. Rev. {\bf D 47} (1993)
3405.}
\lref\sen{A. Sen, ``Non-BPS states and branes in string
theory,'' hep-th/9904207, and references therein.}
\lref\witcs{E. Witten, ``Noncommutative geometry and string field
theory,'' Nucl. Phys. {\bf B 268} (1986) 253.}
\lref\shat{S. Shatashvili, ``Comment on the background independent
open string theory,'' hep-th/9303143, Phys. Lett. {\bf B 311} (1993) 83;
``On the problems with background independence in string theory,''
hep-th/9311177.}
\lref\sfttach{V. A. Kostelecky and S. Samuel, ``The static
tachyon potential in the open bosonic string theory,'' Phys. Lett. {\bf
B 207} (1988) 169;
``On a nonperturbative vacuum for the open bosonic string,''
Nucl. Phys. {\bf B 336} (1990) 263;
A. Sen and B. Zwiebach, ``Tachyon condensation in string
field theory,'' hep-th/9912249, JHEP {\bf 0003} (2000) 002;
N. Moeller and W. Taylor, ``Level truncation and
the tachyon in open bosonic string field theory,'' hep-th/0002237,
Nucl. Phys. {\bf B 583} (2000) 105.}
\lref\lumps{J.A. Harvey and P. Kraus, ``D-branes as unstable
lumps in bosonic open string theory,'' hep-th/0002117, JHEP {\bf 0004}
(2000) 012; R. de Mello Koch, A. Jevicki, M. Mihailescu, and R. Tatar,
``Lumps and $p$-branes in open string field theory,'' hep-th/0003031,
Phys. Lett. {\bf B 482} (2000) 249;
N. Moeller, A. Sen and B. Zwiebach, ``D-branes as tachyon
lumps in string field theory,'' hep-th/0005036, JHEP {\bf 0008}
(2000) 039.}
\lref\kmm{D. Kutasov, M. Mari\~no, and G. Moore, ``Some
exact results on tachyon condensation in string
field theory,'' hep-th/0009148, JHEP {\bf 0010} (2000) 045.} 
\lref\gs{A. Gerasimov and S. Shatashvili, ``On exact tachyon
potential
in open string field theory,'' hep-th/0009103, JHEP {\bf 0010} (2000) 034.}
\lref\berko{N. Berkovits, ``Super-Poincar\'e invariant
superstring field theory,'' hep-th/9503099,
Nucl. Phys. {\bf B 450} (1995) 90. ``A new approach to 
superstring field theory,'' hep-th/9912121.}
\lref\ltsuper{N. Berkovits, ``The tachyon potential in
open Neveu-Schwarz string field theory,'' hep-th/0001084,
JHEP {\bf 0004} (2000) 022; N. Berkovits, A. Sen and
B. Zwiebach, ``Tachyon condensation in superstring field
theory,'' hep-th/0002211, Nucl. Phys. {\bf B 587} (2000) 
147; P. De Smet and J. Raeymaekers,
``Level-four approximation to the tachyon potential
in superstring field theory,'' hep-th/0003220, JHEP
{\bf 0005} (2000) 051; A. Iqbal and
A. Naqvi, ``Tachyon condensation on a non-BPS
D-brane,'' hep-th/0004015.}
\lref\at{O.D. Andreev and A.A. Tseytlin,
``Partition function representation for the open
superstring action,'' Nucl. Phys. {\bf B 311} (1988) 205; 
A.A. Tseytlin, ``Sigma model approach to string theory 
effective actions with tachyons,'' hep-th/0011033.}
\lref\rsz{L. Rastelli, A. Sen and B. Zwiebach, ``String field theory 
around the tachyon vacuum,'' hep-th/0012251. ``Classical solutions in 
string field theory around the tachyon vacuum,'' hep-th/0102112.}
\lref\kl{P. Kraus and F. Larsen, ``Boundary string field theory of the 
D-${\overline {\rm D}}$ system,'' hep-th/0012198.}
\lref\japon{T. Takayanagi, S. Terashima and T. Uesugi, ``Brane-antibrane 
action from boundary string field theory,'' hep-th/0012210.}

\newsec{Introduction}

The problem of open string tachyon 
condensation has 
been studied intensively in the last 
two years. String 
field theory has shown to be a very 
fruitful framework 
to address this problem, and it has 
provided strong evidence 
for Sen's conjectures \sen. There are 
currently two different 
approaches to (bosonic) string field theory: 
the cubic string 
field theory formulated by Witten \witcs, 
and the so-called 
boundary string field theory (BSFT) of 
Witten and Shatashvili \bsft\wittwo\shat. 
In the cubic theory, the approach to 
tachyon condensation has 
relied so far in an approximation scheme 
called level truncation 
\sfttach\lumps \foot{A different approach has been 
proposed recently in \rsz.}. 
On the other hand, BSFT provides exact results 
on the tachyon potential \gs\kmm\ and on D-brane tensions \kmm. 

The bosonic BSFT has a nice geometric formulation 
in terms of the Batalin-Vilkovisky (BV) formalism \bsft. In 
this formulation, the string field action $S$ is defined up 
to a constant by a locally 
exact one-form $dS$ in the space of couplings for boundary 
perturbations. 
To define $dS$, one needs a closed two-form $\omega$ of ghost number 
$-1$, together with 
a vector field $V$ of ghost number $1$ 
which is a symmetry of $\omega$ ({\it i.e.}, ${\cal L}_V 
\omega =0$). When this is the 
case, the string field action is defined by 
\eqn\sfa{
dS = \iota_V \, \omega,}
which is indeed closed (since $d\omega={\cal L}_V \omega=0$), and then 
locally exact. 

In BSFT, the vector field $V$ and two-form are defined as follows 
(when ghosts and matter are decoupled). Take a worldsheet action with 
the structure 
\eqn\wsac{
I=I_{\rm bulk} + \int_{\partial \Sigma} {\cal V},}
where $I_{\rm bulk}$ is the standard closed string background 
for matter and ghost, and ${\cal V}$ is a generic matter operator. 
Consider the ghost-number $1$ operator ${\cal O}= c {\cal V}$. 
We then define the two-form $\omega$ as follows:
\eqn\omegaf{
\omega ( {\cal O}_1, {\cal O}_2 ) ={1 \over 2} \oint d\tau \oint d\tau' 
\langle {\cal O}_1 (\tau) {\cal O}_2 (\tau') \rangle ,}
where the brackets denote correlation functions in the theory with 
worldsheet action \wsac. Since the ghost number of the vacuum is $-3$, 
$\omega$ has indeed ghost number $-1$. The vector field $V$ is defined by the 
BRST operator $Q_{\rm BRST}$ in the bulk theory, which has ghost number $1$, 
and \sfa\ gives:
\eqn\dsromano{
dS = {1 \over 2} \oint d\tau \oint d\tau' \langle d {\cal O}(\tau) 
\{ Q_{\rm BRST}, {\cal O} (\tau')\} \rangle.}
The fact that $d\omega={\cal L}_V \omega=0$ are consequences 
of rotation invariance, and invariance under $b_{-1}$ and 
$Q_{\rm BRST}$. The equation \dsromano\ 
defines the spacetime string field action in BSFT. 
Starting from \dsromano, one can show that $S$ is given 
by \wittwo\shat:
\eqn\sromano{
S=\bigl( \beta^k {\partial \over 
\partial x^k} + 1\bigr) Z,}
where the $x^k$ are couplings for boundary operators, and 
$\beta^k$ are the corresponding $\beta$-functions.

In the superstring case, the level truncation scheme has been also 
developed \ltsuper\ in the superstring field theory formulated by 
Berkovits \berko, as 
well as in some other models \akbm, providing again some 
important evidence for Sen's conjectures. In \kmms, the problem 
of constructing a boundary 
superstring field theory (BSSFT) was addressed. Based on some 
previous works on the sigma model approach to superstring field 
theory \at, it was proposed that the spacetime superstring field 
action is given  
by the (perturbed) partition function on the disc. This is in 
contrast with the bosonic case \sromano, where the string 
field action differs from the partition function in 
a term involving the $\beta$-functions.

In this paper we make a few steps towards a geometric construction  
(a la BV) of BSSFT. We restrict ourselves to 
the NS sector of the open superstring, 
and to perturbations where the ghosts and the matter are decoupled. 
In section 2 we make a proposal for $dS$ which is a simple 
generalization of \dsromano\ to the superstring case. In section 3, 
we make some detailed computations for exactly solvable tachyon 
perturbations, and find that $S=Z$. In section 4, we give a simple 
and  general argument to show that $S=Z$. This gives further support 
to the proposal made in \kmms\ from a different, more geometric 
point of view. 
We end in section 5 with some open problems.

\newsec{Proposal for BSSFT}
 
In order to find a BV formulation of BSSFT, which generalizes the 
construction in \bsft, we have to construct a closed one-form in the 
space of coupling constants. We will restrict ourselves to the 
NS sector of the open superstring. We take as the perturbed action on the disc,
\eqn\pert{
I=I_{\rm bulk} + I_{\Gamma} + \int_{\partial \Sigma}
G_{-1/2} {\cal O}.}
In this equation, $I_{\rm bulk}$ is the RNS bulk action
including the $bc$ and the $\beta \gamma$ (super)ghosts.
$I_{\Gamma}$ is the action for the auxiliary fermionic
superfield living in the boundary \hkm\kmm:
\eqn\bfer{
I_{\Gamma}=\oint {d \tau \over 2 \pi} d\theta \Gamma D \Gamma,}
where
\eqn\ga{
\Gamma =\mu + \theta F}
and $D=\partial_{\theta} + \theta \partial_{\tau}$.
This auxiliary superfield is needed in order to incorporate the 
GSO$(-)$ states. Finally, ${\cal O}$ is the lowest component of a worldsheet superfield:
\eqn\super{
\Psi ={\cal O} + \theta G_{-1/2} {\cal O},}
{\it i.e.} $G_{-1/2} {\cal O}$ is a NS matter operator in the
0-picture.
 
A natural proposal for the superstring field action is:
\eqn\bssft{
dS = {1 \over 8} 
\oint {d\tau d\tau' \over (2\pi)^2} \langle  c(\tau) d {\cal V}^{(-1)}(\tau)
\{ Q_{\rm BRST}, c(\tau') {\cal V}^{(-1)}(\tau') \} \rangle,}
where
\eqn\oper{
{\cal V}^{(-1)}={\rm e}^{-\phi} {\cal O}}
is the usual vertex in the $(-1)$-picture, and the correlation
function is computed in the theory with action \pert. The operator
$Q_{\rm BRST}$ is the BRST operator of the NSR string \fms, and
we assume that it does not act on the auxiliary
field $\Gamma$. This is a natural assumption, since
$\Gamma$ comes from Chan-Paton degrees of freedom, and lives
in the boundary, while $Q_{\rm BRST}$ comes from the bulk CFT.
Notice that we have the appropriate number of (super)ghost insertions
to soak up all the zero modes, and that we are 
working in the ``small" Hilbert space, as in the formulation given in
\witten.

The above proposal is a natural generalization of
\dsromano. The vector field in the space of couplings is 
again given by the BRST operator $Q_{\rm BRST}$, and the two-form 
$\omega$ is a natural generalization of \omegaf:
\eqn\newomega{
\omega ( {\cal O}_1, {\cal O}_2 ) = {1 \over 8} \oint {d\tau  d\tau'
\over (2\pi)^2} 
\langle c(\tau) {\rm e}^{-\phi(\tau)}{\cal O}_1 (\tau) 
c(\tau') {\rm e}^{-\phi(\tau')}{\cal O}_2 (\tau') \rangle ,}
after taking into account 
the superconformal ghosts. Since ${\rm e}^{-\phi}$ has zero ghost number, 
$\omega$ has again ghost number $-1$. It follows from the 
arguments in \bsft\ that $dS$ is a closed one-form.

\newsec{Some explicit computations}

In order to illustrate the above proposal, we will compute 
$dS$ explicitly when the boundary perturbation is a constant 
or a linear tachyon field. As it has been shown it \kmms, in these 
cases the theory is exactly solvable (since it is a free theory).  
The superfield describing a tachyon
profile is \hkm\kmms:
\eqn\supertach{
\Psi =\Gamma \, T({\bf X}),}
therefore
\eqn\lowest{
{\cal O} =\mu \, T(X),}
and
\eqn\highest{
G_{-1/2} {\cal O} =F \, T(X) + \psi^{\rho}\mu \partial_{\rho}T.} 
The action for the auxiliary superfield and the boundary perturbation 
add up to \kmms:
\eqn\bound{
\oint {d\tau \over 2\pi} \bigl( F^2 + \dot \mu \mu + FT(X) +
 \psi^{\rho}\mu \partial_{\rho}T \bigr).}
 
\subsec{Constant tachyon}
 
The computation for the constant tachyon is particularly simple. 
First, we have
\eqn\brst{
\{ Q_{\rm BRST}, c \mu T(X) \} ={1 \over 2} c \partial c {\rm e}^{-\phi}
\mu T(X) + 2 c \partial c {\rm e}^{-\phi}\mu  
{\partial ^2 T \over \partial X^2}.}
Since $T$ is a constant, we find that
\eqn\dsconst{
dS=-{1 \over 16}\oint {d\tau d\tau'\over (2\pi)^2} \langle c \partial c (\tau)
c (\tau')\rangle \langle {\rm e}^{-\phi}(\tau)
{\rm e}^{-\phi}(\tau')\rangle \langle \mu(\tau) \mu (\tau') \rangle 
 T \, d T \exp \bigl\{ -{1 \over 4} T^2 \bigr\}.}
In this equation, the correlators are evaluated in the decoupled $bc$,
$\phi$ and $\mu$ theories, and we have integrated out the auxiliary field
$F$ as in \kmm. Using now the correlation functions,
\eqn\correl{
\eqalign{
\langle c \partial c (\tau) c (\tau')\rangle =& -4\, 
\sin^2 \biggl( {\tau -\tau'\over 2} 
\biggr),\cr
\langle {\rm e}^{-\phi}(\tau) {\rm e}^{-\phi}(\tau')\rangle = 
& -{ 2 \over \sin \bigl( {\tau -\tau'\over 2} \bigr)},\cr
\langle \mu(\tau) \mu (\tau') \rangle = & {\pi \over 2} 
\epsilon (\tau -\tau')
.\cr
}
}
one obtains:
\eqn\finalcons{
dS =  d \bigl( {\rm e}^{-{1 \over 4} T^2}\bigr).}
Integrating \finalcons, we find that $S=Z$ (we have not included in 
\finalcons\ a global normalization proportional to the volume of the 
D9-brane).
 
\subsec{Linear tachyon}
 
The computation of $dS$ for a linear tachyon
\eqn\linear{
T(X)=u X}
is more involved.
The definition gives,
\eqn\dslin{
dS=-{1 \over 16} 
\int {d\tau d\tau'\over (2\pi)^2} \langle c \partial c (\tau)
c (\tau')\rangle \langle {\rm e}^{-\phi}(\tau)
{\rm e}^{-\phi}(\tau')\rangle \langle \mu(\tau) \mu (\tau') \rangle 
 u du \langle X(\tau) X (\tau') \rangle
.}
Now the action \bound\ for $\mu$ is a Gaussian with a linear term. 
We then shift,
\eqn\shift{
\tilde \mu=\mu +{1 \over 2} {1 \over \partial_{\tau}} \psi^{\rho}
 \partial_{\rho} T, }
where
\eqn\inverse{
{1 \over \partial_{\tau}} f(\tau) ={1 \over 2} 
\int d\tau' \epsilon (\tau-\tau')
f(\tau').}
Using all this, \dslin\ becomes the sum of two terms. The first
one is simply,
\eqn\first{
dS_1 =-{\pi\over 8}  dy Z(y) \int {d\tau d\tau'\over (2\pi)^2} \sin
\biggl( {\tau -\tau'\over 2} \biggr) \epsilon (\tau-\tau')
G_B (\tau-\tau',y),}
where $y=u^2$, $Z(y)$ is the partition function, and
$G_B (\tau-\tau',y)$ is the perturbed bosonic
propagator \wittwo,
\eqn\bprop{
G_B(\tau-\tau',y)= {2 } \sum_{k \in {\bf Z}} {1 \over |k| + y}
{\rm e}^{i k (\tau-\tau')}.}
The second piece is:
\eqn\second{
dS_2 = -{ 1\over 64} y dy Z(y) \int {d\tau d\tau'\over (2\pi)^2} d\sigma
d\sigma'
\sin \biggl( {\tau -\tau'\over 2} \biggr) \epsilon (\tau-\sigma)
\epsilon (\tau'-\sigma')G_B (\tau-\tau',y)
G_F (\sigma-\sigma',y),}
where \kmm\
\eqn\fprop{
G_F (\tau-\tau',y)={2i} \sum_{k \in {\bf Z} + {1\over 2}}
{k \over |k| + y} {\rm e}^{i k (\tau-\tau')}}
is the perturbed fermionic propagator.
Using the explicit representation of the step function as
\eqn\step{
\epsilon (\tau-\tau')={2 \over \pi} \sum_{k \in {\bf Z} + {1\over 2}>0}
{\sin k(\tau-\tau')\over r},}
we find for the first term
\eqn\seriesone{
dS_1 = {1 \over 4}  dy Z(y) \biggl\{ -\sum_{n=0}^{\infty}
{1 \over (n+y)(n+1/2)} +\sum_{n=1}^{\infty}
{1 \over (n+y)(n-1/2)}\biggr\},}
and for the second term,
\eqn\seriestwo{
\eqalign{
dS_2 = &{1 \over 4} y dy Z(y) \biggl\{ \sum_{n=0}^{\infty}
{1 \over (n+y)(n+y+1/2)(n+1/2)} \cr 
&\,\,\,\,\,\,\,\,\ -\sum_{n=1}^{\infty}
{1 \over (n+y)(n+y-1/2)(n-1/2)}\biggr\}.\cr}}
Therefore, the one-form \bssft\ is in this case:
\eqn\total{
dS = -{1 \over 4} dy Z(y)\biggr\{ \sum_{n=0}^{\infty}
{1 \over (n+y)(n+y+1/2)} -\sum_{n=1}^{\infty}
{1 \over (n+y)(n+y-1/2)}\biggr\}.}
Notice that all the infinite sums involved here are 
convergent. We can explicitly evaluate 
them by decomposing in simple fractions and using the formula
\eqn\psidif{
\psi(x)-\psi(y)=\sum_{n=0}^{\infty}\biggl( {1 \over 
y+n}-{1 \over x+n}\biggr),} where 
$\psi(x)$ is the logarithmic derivative of the 
$\Gamma$ function, as well as the doubling formula
\eqn\double{
\psi(2x)=\log 2 +{1 \over 2} \bigl( \psi(x) + \psi (x+1/2) \bigr).}
The infinite sums in \total\ add up to:
\eqn\infi{
2\bigl\{ -4 \log 2 -{1 \over y} -4 \bigl(\psi(y)-\psi(2y)\bigr)\bigr\}.} 
Using the explicit results of \kmms, it is easy to see that this is 
nothing but the correlator 
\eqn\reg{
\langle X^2 + \psi {1 \over \partial_\tau}\psi\rangle}
regularized with a point-splitting procedure which preserves 
supersymmetry.
We then find: 
\eqn\finallin{
dS = -{1 \over 4}\langle X^2 + \psi {1 \over \partial_\tau}\psi\rangle Z (y) dy
,}
On the other hand, it was shown in \kmms\ that
\eqn\diff{
{d \log Z \over d y} =-{1 \over 4}
\langle X^2 + \psi {1 \over \partial_\tau}\psi\rangle
.}
Therefore, we can integrate $dS$ to obtain
\eqn\finaltwo{
S =Z(y),}
up to an additive constant. It is interesting to note that 
the convergent sums involved in $dS$ give automatically the 
supersymmetric regularization 
of the propagators proposed in \kmms.

\newsec{$S=Z$}
The explicit computations of the previous section suggest 
that the action defined in \bssft\ is in fact the partition function on the 
disc, {\it i.e.} $S=Z$. In this section we give a general 
argument showing that this is in fact the case, at least in 
the case in which matter and ghosts 
are decoupled. To do this, we adapt the argument 
given in \wittwo. The idea is to consider two decoupled subsystems, 
with partition functions $Z_1$ and $Z_2$, in such a way that the 
combined partition function is $Z=Z_1 \, Z_2$. Let us now 
evaluate $dS$ in the case in which we turn on a series of operators 
in the GSO$(-)$ sector. If ${\cal O}_i$ denotes a basis of 
operators for the first subsystem, and ${\widetilde {\cal O}}_j$ a 
basis of operators in the second subsystem, then the general operator 
${\cal O}$ has the form:
\eqn\genop{
{\cal O} =  \sum_i x^i {\cal O}_i 
+ \sum_j y^j {\widetilde {\cal O}}_j.}
In this equation, $x^i$ and $y^j$ are coupling constants for the 
first and second subsystems, respectively, and the operators in the 
GSO$(-)$ sector have the form, 
\eqn\opers{
{\cal O}_i=\mu_1 {\cal B}_i, \,\, {\widetilde {\cal O}}_j=\mu_2 
{\widetilde {\cal B}}_j,}
where $\mu_1$, $\mu_2$ are 
the boundary fermions for the first and second subystems. and 
${\cal B}_i$, ${\widetilde {\cal B}}_j$ are bosonic matter operators for the 
first and second subsystems. The key fact 
is that, since the two subsystems are decoupled, 
\eqn\dec{
\langle \mu_1 (\tau) \, \mu_2 (\tau') \rangle =0.}
By assumption, the BRST charge does not act on the boundary fermions, 
and after evaluating the ghost correlation functions, we find:
\eqn\dsfirst{
\eqalign{
dS= & \oint d\tau d\tau'  \langle d{\cal O}(\tau) \biggl[ 
\bigl(\sum_i A^i (x, \tau-\tau') {\cal B}_i (\tau')\bigr) \mu_1 (\tau') 
\cr & \,\,\,\,\,\,\ +
 \bigl(\sum_j {\widetilde A}^j (y, \tau-\tau') {\widetilde {\cal B}}_j 
(\tau')\bigr) \mu_2 (\tau') \biggr] \rangle,\cr}}
where $A^i (x, \tau-\tau')$, ${\widetilde A}^j (y, \tau-\tau')$ 
are some functions for the first and second subsystem, respectively, whose 
detailed structure will not be relevant for the argument. After 
evaluating the correlation functions and performing the integrals, 
we find, using \dec, that $dS$ can be written as
\eqn\strucds{
dS =\bigl( \sum_i dx^i a_i(x) \bigr) Z_2 + 
Z_1 \bigl( \sum_j dy^j {\widetilde a}_j(y)\bigr),}
where again $a_i(x)$, ${\widetilde a}_j(y)$ are some functions 
of the couplings for the first and second subsystems. Notice that, in 
contrast to the situation analyzed in \wittwo, the presence of the 
decoupled boundary fermions implies that correlation functions of 
the form $\langle {\cal O}_i {\widetilde {\cal O}}_j \rangle$ vanish and 
do not appear in the final expression for $dS$. 

We now use the key fact that $dS$ is a closed one-form, {\it i.e.} 
$d^2 S=0$. Setting to zero the coefficient of the $dx^i \wedge dy^j$ 
terms gives
\eqn\escero{
-\bigl( \sum_i dx^i a_i(x) \bigr) dZ_2 + 
dZ_1 \bigl( \sum_j dy^j {\widetilde a}_j(y)\bigr)=0.} 
Since $dZ_{1,2}$ only depends on $x$ (respectively, $y$), we 
find that 
\eqn\resul{
g dZ_1=  \sum_i dx^i a_i (x), \,\,\,\ g dZ_2=  \sum_j dy^j 
{\widetilde a}_j (y),}
for some constant $g$. This fixes the unknown functions $a_i(x)$, 
${\widetilde a}_j (y)$ in terms of the partition functions of the 
subsystems. In particular, one finds that, up to an additive constant, 
\eqn\finalres{
S=g \, Z_1 \, Z_2.}
The constant $g$ can be fixed by computing $S$ for some particular 
couplings, as we did before. We therefore find $g=1$ and 
finally 
\eqn\finalres{
S=Z.}
This argument can be extended to perturbations in the GSO$(+)$ 
sector. The key ingredient is that the operator 
${\cal O}$ is again fermionic. Since the two 
subsystems are decoupled, 
correlation functions of the form $\langle {\cal O}_i {\widetilde 
{\cal O}}_j\rangle$ vanish, and the above argument goes through. This 
shows that $S=Z$ for operators in the NS sector, 
when matter and ghosts are decoupled.

\newsec{Open problems}
In this note we have made a proposal for a BV 
formulation of BSSFT, and we have shown that the spacetime superstring 
field action turns out to be the partition function on the disc, in 
agreement with \kmms. There are many problems related to this formulation 
that have not been addressed. Some of these 
problems are the following:

1) As noticed in \shat\kmm, the BV formulation of BSFT is not 
covariant. The covariant definition of the string field action 
is rather, 
\eqn\cov{
{\partial S \over \partial x^i} =\beta^j {\cal G}_{ij},}
where ${\cal G}_{ij}$ is a positive-definite metric. It would 
be very interesting to see if our definition admits such a 
covariant extension, involving a metric which is manifestly 
positive-definite. 
This would immediately show that the superstring field action 
is monotonically decreasing along RG flows, as in the bosonic 
case.  

2) Here we have concentrated on the non-projected 
open NSR superstring, which describes the non-BPS D9-brane of 
type IIA theory. It would 
be interesting to extend our considerations to the D-${\overline {\rm D}}$ 
system, which has been analyzed from the point of view of BSFT in 
\kl\japon. 

3) An important open problem is the inclusion of
the R sector in this formulation of BSSFT. This is a problematic 
issue in the existing formulations 
of superstring field theory, and BSSFT may turn out to be a better 
framework.  

\bigskip
\noindent{\bf Acknowledgements:}
I would like to thank Nathan Berkovits and Greg Moore 
for collaboration at the initial stage of this work, and 
for many clarifying discussions. I am also grateful to 
Hong Liu and Greg Moore for their comments on the final 
version of the paper. Finally, I would like to thank the 
Departamento de F\'\i sica de Part\'\i culas da Universidade 
de Santiago de Compostela, for hospitality. This work has been 
supported by DOE grant DE-FG02-96ER40949. 

\listrefs
\bye